\documentclass[%
reprint,
showpacs,preprintnumbers,
nofootinbib,
amsmath,amssymb,
aps,
prd,
]{revtex4-1}

\usepackage{graphicx}

\usepackage[dvips]{color}

\usepackage{tikz}

\usepackage{ulem}

\newcommand{\ene}[1]{E_{{#1}}}
\newcommand{\nobf}[1]{{#1}}

\begin{document}

\title{Form factor and width of a quantum string}
\author{Arttu Rajantie}
\email{a.rajantie@imperial.ac.uk}
\affiliation{
Department of Physics,
Imperial College London SW7 2AZ,
U.K.}
\author{Kari Rummukainen}
\email{kari.rummukainen@helsinki.fi}
\author{David J. Weir}
\email{david.weir@helsinki.fi}
\affiliation{
Department of Physics and Helsinki Institute of Physics,
PL 64 (Gustaf H\"{a}llstr\"{o}min katu 2),
FI-00014 University of Helsinki,
Finland
}
\date{January 8, 2013}
\begin{abstract}
In the Yang-Mills theory, the apparent thickness of the confining string is known to grow logarithmically when its length increases. The same logarithmic broadening also happens to strings in other quantum field theories and domain walls in statistical physics models. Even in quantum field theories, the correlators used to measure and characterise this phenomenon are analogous to those in statistical physics. In this paper we describe it using the string form factor which is a meaningful quantum observable, obtainable in principle from scattering experiments. We show how the form factor can be obtained from field correlation functions calculated in lattice Monte Carlo simulations. We apply this method to 2+1-dimensional scalar theory in the strong coupling limit, where it is equivalent to the 3D Ising model, and through duality also to 2+1-dimensional $\mathbb{Z}_2$ gauge theory. We measure the string form factor by simulating the Ising model, and demonstrate that it displays the same logarithmic broadening as observed by other quantities.
\end{abstract}
\preprint{Imperial/TP/2012/AR/4}
\preprint{HIP-2012-22/TH}
\maketitle

\section{Introduction}
\label{sec:introduction}

String-like excitations play an important role in many quantum field theories, for example the confining string in Yang-Mills theory~\cite{'tHooft:1973jz} and cosmic strings in some Grand Unified Theories~\cite{Kibble:1976sj}. String excitations are also closely related to the physics of interfaces in statistical physics~\cite{Buff:1965zz,Jasnow1984}. 

In the semiclassical approximation, the string is described by a solution of the field equations which is time-independent and translation invariant along the string, and describes the properties of the string on all length scales. Quantum mechanically, the picture is very different because the string carries massless Goldstone modes whose quantum fluctuations dominate the dynamics on small scales. For example, the string width appears to depend logarithmically on its length because of these fluctuations.

Because the behaviour of the Goldstone modes is independent of microscopic details, this logarithmic broadening is a general property of any strings irrespective of the theory, and it has been studied extensively in Yang-Mills theory and spin models, both analytically~\cite{Luscher:1980iy,Caselle:2006wr,Gliozzi:2010zt,Caselle:2012rp,Meyer:2010tw}, and numerically~\cite{BuerknerStauffer,MonLandauStauffer,Hasenbusch:1991sq,Hasenbusch:1992zz,Caselle:1995fh,Muller:2004vv,Gliozzi:2010zv,Gliozzi:2010jh}.
In the context of the confining string in QCD or Yang-Mills theory, it is generally seen as a problem, because it makes it impossible to measure ``intrinsic'' properties of the string which one would like to know for phenomenological descriptions of confinement~\cite{Kopf:2008hr,Caselle:2012rp}.

The previous numerical studies of strings in quantum field theory have typically measured correlators of plaquettes near the string, probing the field in the vicinity of the string. However, if we truly had access to a real, physical string, we would probably probe it experimentally by scattering particles off it. This motivates us to focus on the string form factor, which is related to the corresponding scattering amplitude and is a well-defined quantum observable.
Building on previous work on kinks and monopoles~\cite{Rajantie:2010fw,Rajantie:2011nq}, we show how the form factor of a string-like topological soliton can be calculated in Monte Carlo simulations.

Because quantum mechanical strings in Euclidean spacetime are equivalent to domain walls in statistical mechanics, we apply the method in practice to calculate the domain wall form factor in the three-dimensional Ising model near the critical point. This theory is in the same universality class as the 2+1-dimensional real scalar field theory and has therefore the same critical behaviour. It is also exactly dual to the confining three-dimensional $\mathbb{Z}_2$ gauge theory, so our conclusions should also be valid for confining strings, at least qualitatively.

\section{String Solution in Scalar Theory}
\label{sec:dwist}
Let us start by considering the 2+1-dimensional real scalar field theory with the Lagrangian
\begin{equation}
 {\cal L}=\frac{1}{2}\partial_\mu\phi \partial^\mu\phi-\frac{1}{2}m^2\phi^2-\frac{1}{4}\lambda\phi^4,
\end{equation}
where $\mu\in\{0,1,2\}$. In the broken phase, which semiclassically corresponds to $m^2<0$, the theory has two vacua
\begin{equation}
 \phi=\pm v=\pm\sqrt{\frac{|m^2|}{\lambda}}.
\end{equation}
Classically, there is also a topologically stable solution
\begin{equation}
\label{equ:dwsol}
 \phi(t,x,y)=v\tanh\left(\frac{m(y-y_0)}{\sqrt{2}}\right),
\end{equation}
which we have chosen to be perpendicular to the $y$ direction.
In 2+1 dimensions, this solution can be thought of either as a string (because it is a one-dimensional object on any time slice) or a domain wall (because it divides spacetime into two pieces). Since we are principally interested in using these objects to model strings in 3+1-dimensional theories, we will refer to it as a string when we discuss it in the context of the 2+1-dimensional theory.

There is a general result~\cite{Buff:1965zz}, applicable to any string-like objects, 
that fluctuations broaden the string so that, whatever its initial shape, its width $w$ is given by
\begin{equation}
\label{eq:logwidth}
w^2 = \frac{1}{2\pi\sigma} \ln \frac{L}{c\xi}
\end{equation}
where $\sigma$ is tension of the string.

However, the arguments used to demonstrate this are based on the dependence of the expectation value of the energy density on the transverse position relative to the string, or other similar quantities which are sensitive to the fluctuations of the string position. There has, therefore, been discussion on whether there is some other observable which would measure the ``intrinsic width'' of the string~\cite{Meyer:2010tw,Caselle:2012rp}. 

\section{String Form Factor}

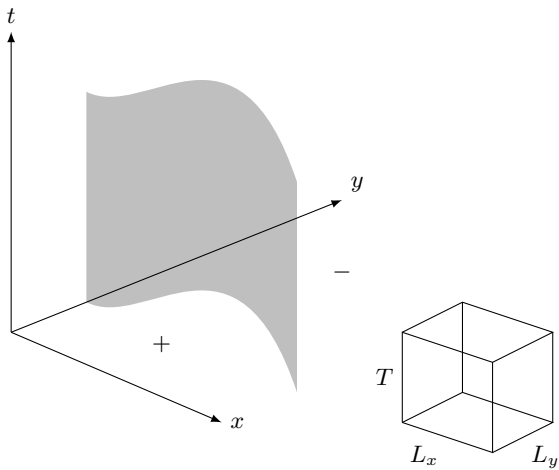
\begin{figure}
\begin{tikzpicture}[scale=0.4]

\fill[lightgray] (2.5,8) -- (2.5,1) .. controls (4.5,0) and (7.5,4) ..(9.5,-2) -- (9.5,5) .. controls (7.5,11) and (4.5,7) .. (2.5,8);
\draw[-latex] (0,0) -- (7,-3) node[right] {$x$};
\draw[-latex] (0,0) -- (11,4.4) node[above right] {$y$};
\draw[-latex] (0,0) -- (0,10) node[above] {$t$};
\draw (5,-0.4) node {$+$};
\draw (11,2) node {$-$};

\draw (13,-3) -- (16,-4) -- (16,-1) -- (13,0) -- (13,-3);
\draw (13,-3) -- (15,-2) -- (18,-3) -- (16,-4);
\draw (13,0) -- (15,1) -- (18,0) -- (16,-1);
\draw (15,-2) -- (15,1);
\draw (18,-3) -- (18,0);

\draw (13,-1.5) node[left] {$T$};
\draw (14.5,-3.5) node[below left] {$L_x$};
\draw (17,-3.5) node[below right] {$L_y$};

\end{tikzpicture}
\caption{\label{fig:setup} System setup. The coordinate $x$ runs along the domain wall (shaded gray) on a timeslice, the twisted boundary conditions are in the $y$ direction and $t$ parameterises the direction along which correlators are measured.}
\end{figure}

\subsection{Definition}
\label{sec:formfactor}

A stable quantum state corresponding to Eq.~(\ref{equ:dwsol}) also exists in the quantum theory. We can define it as the lowest topologically non-trivial energy eigenstate, and we denote it by $|0\rangle$. 
To be precise, there is a degenerate set of such states corresponding to different orientations of the string, and we again choose the string that is oriented along the $x$ direction as shown in Figure~\ref{fig:setup}.

The string ground state $|0\rangle$ has zero momentum and, in the infinite-volume limit, its energy is $E_0=\sigma L$, where the constant $\sigma$ is the string tension. We can obtain moving string states by boosting the ground state $|0\rangle$ in the $y$ direction. We denote these states by $|p\rangle$, where $p$ is the momentum of the string in $y$ direction. The energy of such a state is $E_{p}=\sqrt{p^2+E_0^2}$.
We normalise these states as
\begin{equation}
\label{equ:statenormalisation}
\langle p'|p \rangle=2\pi \delta(p' -p).
\end{equation}
This normalisation is not Lorentz invariant under relativistic boosts in the $y$ direction, but because the string has a linearly divergent energy, we can safely restrict ourselves to only non-relativistic motion.

For any local operator $\hat{\mathcal{O}}$, such as the field operator $\hat\phi$, we can now define the corresponding form factor as
\begin{equation}
f(p_2,p_1) = \langle p_2 | \hat{\mathcal{O}}(0) | p_1 \rangle,
\end{equation}
where by $\hat{\mathcal{O}}(0)$ we mean the operator in coordinate space at position $(x,y)=(0,0)$.

\subsection{Semiclassical Limit}
\label{sec:semiclassical}
In the semiclassical limit~\cite{Goldstone:1974gf}, the form factor is given by the Fourier transform of the classical profile $\mathcal{O}_\text{cl}(y)$ of the quantity $\mathcal{O}$. Taking matrix elements of the Heisenberg equation of motion for the classical field we find
\begin{eqnarray}
\label{equ:classff}
f(p_2,p_1) &=& \langle p_2 | \hat{\mathcal{O}}(0) | p_1 \rangle \nonumber\\ 
&=& \int \mathrm{d} y \; e^{i(p_2 - p_1)y} \mathcal{O}_\text{cl} (y)\nonumber\\
&=& \tilde{O}_\text{cl} (p_2 - p_1).\nonumber\\
\end{eqnarray}
This expression is valid only in the non-relativistic limit, when $|p_1|,|p_2|\ll \sigma L$. In this limit the form factor is a function of the momentum difference $k\equiv p_2-p_1$ only, as a direct consequence of the Galilean invariance, so we will denote it by $f(k)$. Although our studies of kinks and monopoles used fully relativistic semiclassical expressions for the form factor, this is a reasonable approximation in the current paper since any results obtained with a string moving relativistically are most probably due to finite-size effects (the amount of energy required to accelerate a string to relativistic velocities on macroscopic scales is too great; the finite box size does not correctly capture this).

In particular, choosing $\hat{\mathcal{O}}=\hat\phi$, we find from Eq.~(\ref{equ:dwsol}) the semiclassical result for the string form factor
\begin{equation}
\label{eq:ffsemiclass}
f(k) = \frac{i\sqrt{2}\pi v}{m\sinh(k\pi/\sqrt{2}m)}.
\end{equation}
At low momenta $k\ll m$, this behaves asymptotically as
\begin{equation}
\label{eq:ftstepfn}
f(k) = v \frac{2 i}{k},
\end{equation}
which corresponds to an infinitesimally thin string
\begin{equation}
\label{eq:stepfn}
\phi_\text{cl}(y) = v \, \mathrm{Sign}(y).
\end{equation}
Because any string would look like this when viewed from long distances, we expect
the low-$k$ asymptotic behaviour (\ref{eq:ftstepfn}) to be valid for any string, irrespective of microscopic details and even in quantum theory. 
Therefore it serves as a useful benchmark for our calculations.

\subsection{Correlator and form factor}
\label{sec:correlators}

It was shown previously in Ref.~\cite{Rajantie:2010fw} how the form factors of point-like solitons such as kinks and monopoles can be calculated from the field correlation function. The same technique works also for strings.

We work in momentum space, taking the Fourier transform in space but not in the time direction. After Wick rotation to Euclidean time,
the two-point correlator of the operator $\hat{\mathcal O}$ has the spectral expansion.
\begin{multline}
\label{equ:spectral}
\langle \mathcal{O}(0,k_x,k_y) \mathcal{O}(t,q_x,q_y)\rangle \\ =\sum_\alpha
\frac{\langle 0| \hat{\mathcal{O}}(k_x,k_y)|\alpha\rangle\langle\alpha| \hat{\mathcal{O}}(q_x,q_y)|0\rangle}{\langle0|0\rangle}
e^{-t(E_\alpha-E_0)},
\end{multline}
where $E_0$ is the energy of the single string ground state. 

A small added complication is that the spacetime is necessarily finite in actual Monte Carlo simulations. We have periodic boundary conditions in the $t$-direction (as well as the $x$-direction along the string), and twisted boundary conditions in the $y$-direction. This periodicity leads us to write the correlator as
\begin{multline}
\label{eq:bigtrace}
\langle \mathcal{O}(0,k_x,k_y) \mathcal{O}(t,q_x,q_y)\rangle \\= \frac{1}{Z} \text{Tr}\; \hat{U}(T-t) \hat{\mathcal{O}}(q_x,q_y) \hat{U}(t) \hat{\mathcal{O}}(k_x,k_y)
\\
=\frac{1}{Z}\sum_{\alpha,\alpha'}\langle \alpha'|\hat{\mathcal{O}}(q_x,q_y)|\alpha\rangle
\langle\alpha|\hat{\mathcal{O}}(k_x,k_y)|\alpha'\rangle e^{-E_{\alpha'}(T-t)-E_\alpha t},
\end{multline}
where $\hat{U}(t)=\exp(-\hat{H} t)$ is the Euclidean time evolution operator, and $Z= \mathrm{Tr}\;\hat{U}(T)$ the partition function for the worldsheet, not the full $Z_\text{tw}$ of the field theory with twisted boundary conditions. In other words, only when $T\to\infty$ is the situation of Eq.~(\ref{equ:spectral}) realised; otherwise the string's worldline is restricted.

With twisted boundary conditions, the states $|\alpha\rangle$, $|\alpha'\rangle$ must have an odd number of strings (in practice, exactly one due to heavy volume suppression factors $e^{-2 \sigma L T}$ in the partition function). Because of momentum conservation they must also have opposite overall momentum $(k_x,k_y)=-(q_x,q_y)$. 
If the momentum is in the $y$ direction, i.e., $k_x=0$, then the lowest such state is the moving string state $|k_y\rangle$ with momentum $k_y$, as defined in Section~\ref{sec:formfactor}. It has energy
\begin{equation}
\ene{k_y}=\sqrt{k_y^2+E_0^2}\approx E_0 + \frac{k_y^2}{2E_0}
\end{equation}
where $E_0$ is the ground state energy of the string.

In contrast, if the momentum has a non-zero $x$ component, $k_x\ne 0$, then the interpolating states must be excited states with some excitation carrying the momentum in the $x$ direction. The lightest such states are massless Goldstone modes which exist because the position of the string breaks translation invariance in the $y$ direction. Because we are only interested in the unexcited states $|k\rangle$, we do not consider this possibility, and instead we simply restrict ourselves to $k_x=0$ from now on. To simplify our notation, we therefore suppress the unneeded argument $k_x$, and write
\begin{equation}
 \hat{\mathcal O}(k_y)\equiv \hat{\mathcal O}(0,k_y),
\end{equation}
and
\begin{equation}
\langle \mathcal{O}(0,k_y) \mathcal{O}(t,q_y)\rangle\equiv
\langle \mathcal{O}(0,0,k_y) \mathcal{O}(t,0,q_y)\rangle.
\end{equation}

The next states in the spectrum contain pairs of massless Goldstone modes, with opposite quantised momenta 
$k_x=2n\pi/L$. These propagate along the string perpendicular to our chosen direction for time evolution of the system. 
The energy of the lowest such state, with $k_x=2\pi/L$, is $\ene{k_y}+4\pi/L$.
Therefore, if we choose $t\gtrsim L$, these states are strongly suppressed relative to the moving string state in the spectral expansion (\ref{eq:bigtrace}).
Consequently, the problem becomes practically identical to the 1+1-dimensional kink case~\cite{Rajantie:2010fw}, and we can approximate Eq.~(\ref{eq:bigtrace}) by an integral over moving string states $|k_y\rangle$ only,
\begin{widetext}
\begin{eqnarray}
\label{equ:fullres}
\left< \mathcal{O} (0,\nobf{k}_y) \mathcal{O}(t,\nobf{q}_y) \right> &=& 
\frac{1}{Z} \int \frac{d k_y'}{2\pi}\frac{d k_y''}{2\pi}\langle \nobf{k}_y'|\hat{\mathcal{O}}(\nobf{q}_y)|\nobf{k}_y''\rangle\langle\nobf{k}_y''| \hat{\mathcal{O}}(\nobf{k}_y) | \nobf{k}_y'\rangle e^{-\ene{k_y'}(T-t)-\ene{k_y''} t}
\nonumber\\
&=&
\frac{L^2}{Z} 2\pi \delta(\nobf{q}_y+\nobf{k}_y) \int  \frac{d k_y'}{2\pi}
|f(\nobf{k}_y',\nobf{k}_y'-k_y)|^2
e^{-\ene{k_y'}(T-t)-\ene{k_y'-k_y}t},
\end{eqnarray}
\end{widetext}
where we have used the result
\begin{eqnarray}
\langle k_y'|\hat{\mathcal{O}}(q_y)|k_y''\rangle
&=&\int dx\,dy\,e^{iq_yy}\langle k_y'|\hat{\mathcal{O}}(x,y)|k_y''\rangle\nonumber\\
&=&\int dx\,dy\,e^{iq_yy}\,e^{i(k_y'-k_y'')y}\langle k_y'|\hat{\mathcal{O}}(0)|k_y''\rangle
\nonumber\\
&=&2\pi L\delta(q_y+k_y'-k_y'')f(k_y',k_y'').
\end{eqnarray}
The choice $t\gtrsim L$ obviously requires a relatively long lattice in the time direction. The dynamics of the longest-wavelength modes then become essentially one-dimensional.

Similarly, we can write the partition function as
\begin{eqnarray}
\label{equ:denominator}
Z &=& \int\frac{d k_y'}{2\pi}\langle \nobf{k}_y'| \hat{U}(T)| \nobf{k}_y'\rangle=L\int\frac{d k_y'}{2\pi}e^{-\ene{k_y'}T}
\nonumber\\
&\approx& L \int \frac{d k_y}{2\pi} e^{-\left(E_0+\frac{k_y^2}{2E_0}\right) T} = L\sqrt{\frac{E_0}{2\pi T}} e^{-E_0 T}.
\end{eqnarray}
This partition function is the individual contribution to the partition function from string's classical worldsheet.

To calculate the integral (\ref{equ:fullres}), we use the saddle point approximation.
The saddle point $\nobf{k_0}$ is found by minimising the action
\begin{equation}
\label{equ:pointaction}
S(\nobf{k}_y') =   \ene{k_y'} (T-t) + \ene{k_y'-k_y} t
\end{equation}
for given $t$. By approximating the integral by a Gaussian around the saddle point, we obtain
\begin{multline}
\langle \mathcal{O}(0,\nobf{k}_y)\mathcal{O}(t,\nobf{q}_y)\rangle = 
\frac{L^2}{Z}2\pi \delta(\nobf{q}_y+\nobf{k}_y)\times \\
\int  \frac{d k_y'}{2\pi} |f(\nobf{k}_y'-\nobf{k}_y,\nobf{k}_y')|^2 
e^{-S(\nobf{k}_0)-\frac{1}{2}S''(k_0)(\nobf{k}_y'-\nobf{k}_0)^2},
\end{multline}
This is not a Gaussian approximation of the correlation function, nor is it a semiclassical stationary phase calculation; it does, however, allow us to rearrange Eq.~(\ref{equ:fullres}) and solve for $|f(\nobf{k}_y',\nobf{k}_y'-\nobf{k}_y)|^2$ in exchange for imposing minor restrictions on what we can measure. In the limit of large $t$ and $T-t$ that we already require, the Gaussian approaches a delta function and we can calculate the integral
\begin{multline}
\label{equ:finalcorr}
\left< \mathcal{O}(0,\nobf{k}_y) \mathcal{O}(t,\nobf{q}_y) \right>
\\
\approx 2\pi \delta(\nobf{k}_y+\nobf{q}_y)L  \left(\frac{T}{E_0 S''(k_0)}\right)^{1/2} 
|f(\nobf{k}_0,\nobf{k}_0-\nobf{k}_y)|^2 e^{-S(k_0)+E_0T}.
\end{multline}

We can use Eq.~(\ref{equ:finalcorr}) to determine the form factor from the field correlator. 
For given $\nobf{k}_y$ and $t$, we find the saddle point $\nobf{k}_0$ by minimising Eq.~(\ref{equ:pointaction})
,
and we obtain 
\begin{multline}
\label{equ:fresult}
f(\nobf{k}_0, \nobf{k}_0-\nobf{k}_y)=\pm i \sqrt{\left< \mathcal{O}(0,\nobf{k}_y) \mathcal{O}(t,-\nobf{k}_y) \right>} \\ \times \frac{1}{L} \left(\frac{E_0S''(k_0)}{T}\right)^{1/4} e^{(S(k_0)-E_0 T)/2}
\end{multline}
for $\mathcal{O}$ odd. Note that the saddle point $k_0$ still depends on $t$.

The result (\ref{equ:fresult}) should be compared with the corresponding result for kinks given in Eq.~(19) of Ref.~\cite{Rajantie:2010fw} (although we have corrected a typographical error here). In that case, the interesting length scales were comparable to the inverse kink mass. Therefore the motion of the kink was relativistic for the corresponding momenta, and it was natural to express the result in terms of rapidities. In the current case, the interesting momenta $k_y\sim\sqrt{\sigma}$ are much lower than the string mass $\sigma L$, and its motion is therefore highly non-relativistic. We can therefore simplify Eq.~(\ref{equ:fresult}) by taking the non-relativistic limit. In that case, $k_0=(t/T)k_y$ and $S''(k_0)=T/E_0$, and the form factor also becomes a function of the the momentum difference $k_y$ only,
\begin{eqnarray}
\label{equ:fresultNR}
f(k_y)&\equiv&
f(\nobf{k}_0, \nobf{k}_0-\nobf{k}_y)\nonumber\\
&=&\pm i \sqrt{\left< \mathcal{O}(0,\nobf{k}_y) \mathcal{O}(t,-\nobf{k}_y) \right>}  \frac{1}{L}  \exp\left(\frac{k_y^2}{4E_0}\frac{t(T-t)}{T}\right).\nonumber\\
\end{eqnarray}
Correspondingly Eq.~(\ref{equ:finalcorr}) simplifies to
\begin{multline}
\label{equ:finalcorr2}
\left< \mathcal{O}(0,\nobf{k}_y) \mathcal{O}(t,\nobf{q}_y) \right>
\\
\approx 2\pi L\delta(\nobf{k}_y+\nobf{q}_y)|f(k)|^2\exp\left(-\frac{k_y^2}{2E_0}\frac{t(T-t)}{T}\right).
\end{multline}
We can also use this result to determine the ground state energy $E_0$ from the correlator measurements.

\subsection{Linear fluctuations}
\label{sec:fluc}
While the primary purpose of this paper is to measure the form factor fully non-perturbatively using lattice Monte Carlo simulations, it is useful to look at the leading-order quantum effects analytically.
The lowest excitations in the system are massless Goldstone modes on the string worldsheet, which are present because the string breaks translation invariance in the $y$ direction spontaneously.
To calculate their effect, we assume a string that moves in the normal direction $y$ without changing its profile. If the profile of the string is $\phi_\text{c}(y)$, and the position of the string is $y(t,x)$, 
then the field configuration is
\begin{equation}
\phi(t,x,y) = \phi_\text{c}(y-y(t,x)).
\end{equation}
Taking the Fourier transform over $x$ and $y$, we find
\begin{equation}
\phi_\text{c}(t,k_x,k_y) = \tilde{\phi}_\text{c}(k_y) \int dx \, e^{i(k_x x + k_y y(t,x))},
\end{equation}
where $\tilde{\phi}_\text{c}(k_y)$ is the Fourier transform of the classical profile,
\begin{equation}
\tilde{\phi}_c(k_y) = \int dy \, e^{ik_y y} \phi_c(y).
\end{equation}
We can now write the correlation function as
\begin{multline}
\langle \phi(t,k_x,k_y) \phi(t',k_x', k_y')\rangle = \tilde{\phi}_c(k_y) \tilde{\phi}_c(k_y')\\ \times \int dx \, dx' \, e^{i(k_x x + k_x' x')} \left< e^{i(k_y y(t,x) + k_y' y(t',x'))}\right>.
\end{multline}
Writing $\Delta x = x' - x$ and $\Delta t = t' - t$, we have
\begin{multline}
\langle \phi(t,k_x,k_y) \phi(t',k_x', k_y')\rangle = 2\pi \, \delta(k_x +k_x') \tilde{\phi}_c(k_y)\tilde{\phi}_c(k_y') \\ \times \int d \Delta x \, e^{-ik_x \Delta x}\left< e^{i(k_y y(0,0) + k_y'y(\Delta t, \Delta x))} \right>.
\end{multline}
Translation invariance in the $y$ direction gives a delta function also for the $k_y$ component, so we can assume $k_y' = -k_y$ and we have
\begin{multline}
\langle \phi(t,k_x,k_y) \phi(t',k_x', -k_y)\rangle = 2\pi \, \delta(k_x+k_x')|\tilde{\phi}_c(k_y)|^2  \\ \times \int d\Delta x\, e^{-ik_x \Delta x}\left< e^{i(k_y (y(0,0)+  y(\Delta t, \Delta x)))} \right>.
\end{multline}
Assuming that $y(t,x)$ is Gaussian, this is equal to
\begin{multline}
\langle \phi(t,k_x,k_y) \phi(t',k_x', -k_y)\rangle = 2\pi \, \delta(k_x+k_x')|\tilde{\phi}_c(k_y)|^2 \\ \times \int d \Delta x \, e^{-ik_x \Delta x}e^{-\frac{1}{2} k_y^2\langle (y(0,0) - y(\Delta t, \Delta x))^2 \rangle }.
\end{multline}

However, if we consider an elongated lattice with spatial size $L$ and time separation $\Delta t \gg L$, then the correlator is very different. Let us start with the two-point correlator of the field $y$, which is given by the Fourier transform of the propagator
\begin{equation}
\langle y(0,0) y(t,x) \rangle = \frac{1}{\sigma} \int \frac{dk}{2\pi} \int \frac{d\omega}{2\pi} \frac{e^{i(kx+\omega t)}}{k^2 + \omega^2},
\end{equation}
and in finite spatial volume, the integral over $k$ becomes a sum over $n$, so we have
\begin{equation}
\langle y(0,0) y(t,x) \rangle = \frac{1}{\sigma L} \sum_{n=-\infty}^{\infty}\int\frac{d\omega}{2\pi}\frac{e^{i(k_nx+\omega t)}}{k_n^2 + \omega^2}
\end{equation}
with $k_n = 2\pi n/L$. This is obviously IR divergent, but we can use it to write
\begin{equation}
\langle (y(0,0)-y(t,x))^2 \rangle = 2\langle y(0,0)^2\rangle - 2\langle y(0,0) y(t,x) \rangle
\end{equation}
which yields
\begin{equation}
\langle (y(0,0)-y(t,x))^2 \rangle = \frac{1}{\sigma L} \left[ t+2\sum_{n=1}^{\infty} \frac{1-e^{-k_n t} \cos(k_n x)}{k_n} \right].
\end{equation}
For $t\gtrsim L$, the exponentials are all very small and we can approximate
\begin{align}
\langle (y(0,0)-y(t,x))^2 \rangle \approx  &\frac{1}{\sigma L} \left[t +
  2\sum_{n=1}^{\infty} \frac{1}{k_n} \right] \\
= & \frac{1}{\sigma L} \left[ t + \frac{L}{\pi} \sum_{n=1}^{\infty} \frac{1}{n} 
\right].
\end{align}
This diverges in the UV, but a minimum length scale $l$ provides a
cutoff $n < L/ \ell$. Then for $L \gg \ell$ one has
\begin{equation}
\sum_{n=1}^{L/\ell}\frac{1}{n} = \ln \frac{L}{\ell}  + \gamma,
\end{equation}
and we find (for $k_x = 0$),
\begin{multline}
\label{eq:broadenedcorrel}
\langle \phi(t,0,k_y) \phi(t',0,-k_y) \rangle \\ 
= L^2 |\tilde{\phi}_c(k_y)|^2 e^{-\frac{k_y^2}{2\pi\sigma} (\ln (L/\ell) + \gamma)} e^{-(k_y^2 / 2\sigma L) t}.
\end{multline}
Inserting this into Eq.~(\ref{equ:fresultNR}), we find the form factor
\begin{equation}
f(k_y)=\tilde{\phi}_c(k_y) e^{-\frac{k_y^2}{4\pi\sigma} (\ln (L/\ell) + \gamma)}.
\end{equation}
Comparing with Eq.~(\ref{equ:classff}), we can see that the effect of the fluctuations is to suppress the form factor at high momenta $k_y\gtrsim \sqrt{2\pi\sigma\ln (L/\ell)}$.
For large $L$, this washes out any structure the classical string solution has at short distances, and using the asymptotic low-momentum behaviour of the classical solution $\tilde{\phi}_c(k_y)\sim 2iv/k_y$, we find
\begin{equation}
\label{eq:analyticf}
f(k)=\frac{2iv}{k} e^{-\frac{k^2}{4\pi\sigma} (\ln (L/\ell) + \gamma)}.
\end{equation}
If interpret, in line with Eq.~(\ref{equ:classff}), this as the Fourier transform of the quantum-corrected domain wall profile $\phi_{\rm eff}(y)$, we find that in coordinate space the domain wall has broadened and has width
\begin{equation}
\label{eq:flucwidth}
 w^2=\frac{1}{2\pi\sigma}(\ln (L/\ell) + \gamma),
\end{equation}
in perfect agreement with Eq.~(\ref{eq:logwidth}).

\section{Ising Model}
\subsection{The model}

To demonstrate the use of Eq.~(\ref{equ:finalcorr}) we use it to calculate the domain wall form factor non-perturbatively near the critical point. This is an interesting calculation because the theory becomes strongly coupled and therefore perturbation theory is not valid. In practice, we do not actually simulate the scalar field theory but rather the 3D Ising model, which is known to be in the same universality class and which will therefore give identical results near the critical point. From a computational point of view, the Ising model is much more convenient because highly efficient 
numerical algorithms are available to simulate it. 

The worldsheet of the string corresponds to a domain wall in the 3D Ising model. In this section we will therefore use the term domain wall, but it should be understood to refer to the same physical object as the term string elsewhere in the paper.
The 3D Ising model is also exactly dual to the confining 2+1-dimensional $\mathbb{Z}_2$ gauge field theory. This duality maps the domain wall to the worldsheet of the confining string of the gauge theory, and therefore our calculation also describes the properties of the confining string in this somewhat simple gauge theory.

The logarithmic broadening of the domain wall is well known also in the Ising model.
In the rough phase (between $\beta_\mathrm{R}$ and $\beta_\mathrm{C}$), the domain wall width $w$ has a logarithmic divergence with domain wall length $L$~\cite{Buff:1965zz} of the form given in Eq.~(\ref{eq:logwidth}).

The Ising model has Hamiltonian
\begin{equation}
H = - \sum_{\langle \mathbf{x}, \mathbf{x}'\rangle} k_{\mathbf{x},\mathbf{x'}} s(\mathbf{x}) s(\mathbf{x}')
\end{equation}
where $\mathbf{x}$ and $\mathbf{x}'$ refer to position vectors in the 3D Euclidean space, and $\langle \mathbf{x}, \mathbf{x}' \rangle$ denotes that the sum runs over all nearest-neighbour links in the lattice. The spins $s(\mathbf{x})$ take the values $\pm 1$. We can denote the size of each direction $L_x$, $L_y$ and $T$ respectively, but for our own calculations we shall always take $L_x = L_y = L$. It has partition function
\begin{equation}
Z_0 = \sum_{\{s(\mathbf{x}) = \pm 1 \}} \exp(-\beta H ),
\end{equation}
and we impose periodic boundary conditions in all three directions, with $k_{\mathbf{x}, \mathbf{x}'}=1$ for translational invariance. However, to create a domain wall, a twist is introduced so that $k_{\mathbf{x}, \mathbf{x}'} = -1$ in the $y$-direction for one value of $y$.

Various techniques exist to measure the free energy of such an object; almost all depend on measuring the ratio

\begin{equation}
F = -\ln \frac{Z_\text{tw}}{Z_0},
\end{equation}
where $Z_0$ is the partition function on a fully periodic lattice and $Z_\text{tw}$ is the partition function with antiperiodic boundary conditions in the $y$-direction.

The free energy of the domain wall can be obtained as a function of $L_x$ and $T$. For a cubic system with $T=L_x=L$, a perturbative calculation yields~\cite{Caselle:2006dv}
\begin{equation}
F_\text{cubic} = \sigma L^2 - \frac{1}{2} \ln \sigma + G - \frac{1}{4\sigma L^2} + \mathrm{O}\left( (\sigma L^2)^{-2}\right),
\end{equation}
where $G\approx 0.29$, and the domain wall tension $\sigma$ corresponds directly to the string tension in the 2+1-dimensional scalar theory.

\subsection{Real-space width measurements}
\label{sec:previouswidth}

Previous measurements of the width of walls and strings have worked in real space. Here we summarise one of the more successful approaches and note that later developments are generalisations of the same idea to different systems or improvements in numerical technique.

In Ref.~\cite{Swendsen1977}, the domain wall width was studied. The results were based on linear functions of local spin operators; in addition the $t$-direction was not privileged in this study. For these reasons, this calculation cannot be considered a direct analogy of a particle scattering off the domain wall.

Note that the geometry used was rather different to ours. In addition, the discussion here follows our own conventions established in Section~\ref{sec:introduction}. First the total magnetisation for slices parallel to the domain wall was measured,
\begin{equation}
m(y) = \frac{1}{L T} \sum_{t,x} s(t,x,y).
\end{equation}
For each configuration, $m(y)$ was calculated, and the minimum value $y_0$ of $y$ obtained. Then the `normalised magnetisation gradient' is
\begin{equation}
\rho(y) = \frac{1}{2 M} \left| m\left( y - y_0 + 1\right) - m\left( y - y_0  \right) \right|
\end{equation}
where
\begin{equation}
M = \frac{1}{L}\sum_{y} m(y)
\end{equation}
is the average magnetisation. For a very sharp and straight domain wall, $\rho(y)$ will only be nonzero for one value of $y$. In reality, fluctuations will smear the result, even for individual configurations. There may also be bubbles of spin in the bulk, or (for small lattice sizes) other odd numbers of domain wall. The height $h$ can be defined as
\begin{equation}
h = \sum_y y \rho(y),
\end{equation}
and then the variance in this expression gives us the mean squared width $w^2$,
\begin{equation}
w^2 = \left< \sum_y \, \rho(y) y^2 - \left( \sum_y \, \rho(y) y \right)^2 \right>;
\end{equation}
the angle brackets denote an ensemble average. Note that if the shifting procedure does not introduce a systematic bias in $\langle h \rangle$ (say, the value of $y_0$ is chosen randomly when there are multiple degenerate minima of $m(y)$), then we have $\langle h \rangle = 0$ which we shall assume shortly when studying how this estimate of $w^2$ behaves.

Anyway, using this data for various lattice sizes (for example, in Ref.~\cite{Hasenbusch:1992zz} the system under consideration had $L_x = L_t$ and $L_y$ up to $27$, so the width either side of the domain wall was fixed), a fit to the ansatz taking the same form as Eq.~(\ref{eq:logwidth}) can be made.

We assume $\langle h \rangle$ = 0, as discussed. Now
\begin{align}
w^2 & = \left< \sum_y \rho(y) y^2 \right> \\
 & = \left< \sum_y  \frac{y^2}{2LM} \left| \sum_{t,x} ( s(t,x,y) -
s(t,x,y-1 )) \right| \right>.
\end{align}
In other words, this measurement technique is weighted towards long-distance fluctuations of the domain wall.  It does not and cannot directly probe the physics of any intrinsic width, nor does it allow for the possibility of different behaviour being visible at short distances as such effects will be washed out. One way of evading this problem is the rescaling approach adopted in Ref.~\cite{Muller:2004vv}.

The approach discussed above is equally applicable to the confining string in $\mathrm{SU}(N)$ gauge theory, as was first carried out in Ref.~\cite{Caselle:1995fh}. Of course, the endpoints of the string must be fixed as well so there is not even translational invariance in such a calculation.

\section{Results}
\label{sec:results}

We use the standard Ising single-cluster algorithm~\cite{Wolff:1988uh}, as well as Metropolis updates to obtain the results presented here. To show the volume-dependence of the ground state energy we use $\beta=0.226$; however the domain wall is so light here that finite-size effects for the form factor for higher $k$ are severe. Therefore the form factor results are presented for values of $\beta$ that are slightly further into the rough phase.

\subsection{Stationary string energy}

\begin{figure}[t]
\begin{centering}
\includegraphics[trim=0.5cm 0.5cm 0.5cm 0.5cm,clip=true,scale=0.333,angle=270]{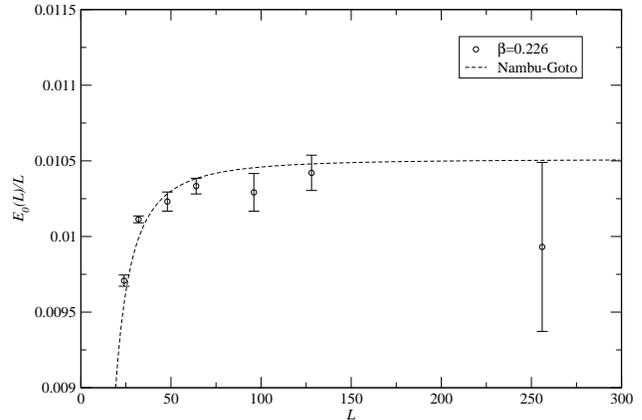}
\caption{\label{fig:tensions} Energy per unit length of the string measured using the correlator with $\beta=0.226$. Units are $a^2$; the error bars are jackknife estimates.}
\end{centering}
\end{figure}

In order to directly recover the string form factor from
correlator measurements, we need an independent estimate of the
string tension $\sigma$. This is obtained from measurements of the the lowest
momentum ($k=\pi/L$) correlator in the presence of the string.

The expression in Eq.~(\ref{equ:finalcorr2}) is used, along with a
contribution from bulk scalar particles; these have the standard
$\cosh$-type long distance behaviour and their contribution to the
correlator can be easily accounted for. To improve fitting
performance, the scalar mass is obtained from fits to correlators with
periodic boundary conditions, leaving only the relative amplitudes of
the two contributions and the surface tension undetermined. This approach is in line with our previous studies along the same lines.

Note that at higher momenta, the bulk
scalar contribution $\sim e^{-\sqrt{m^2 + k^2}t}$ decays much more rapidly with $t$ than the string contribution $\sim e^{-\frac{k^2}{2E_0}t}$, so discarding the first few points at either end of the lattice allows us to ignore this issue when measuring the form factor; Figure~\ref{fig:dwffdist} also shows this effect.

As expected, we see is evidence for the L\"{u}scher term in our string tension measurements. For a closed Nambu-Goto string, the ground state energy is
\begin{equation}
\label{eq:nggs}
E_0(L) = \sigma L \sqrt{1-\frac{\pi}{3 \sigma L^2}}.
\end{equation}
Unfortunately, we cannot rule out the massless worldsheet modes affecting the quality of the data for the smallest of our lattice sizes. The results are shown for $\beta=0.226$ in Figure \ref{fig:tensions}, along with a fit to Eq.~(\ref{eq:nggs}).

Nonetheless, the quality of this fit gives us confidence that it is principally the ground state of the string that gets excited when we carry out measurements at large time separation; this is in line with the assumptions of Section~\ref{sec:correlators}.

\subsection{String form factor}

\begin{figure}[t]
\begin{centering}
\includegraphics[trim=0.5cm 0.5cm 0.5cm 0.5cm,clip=true,scale=0.333,angle=270]{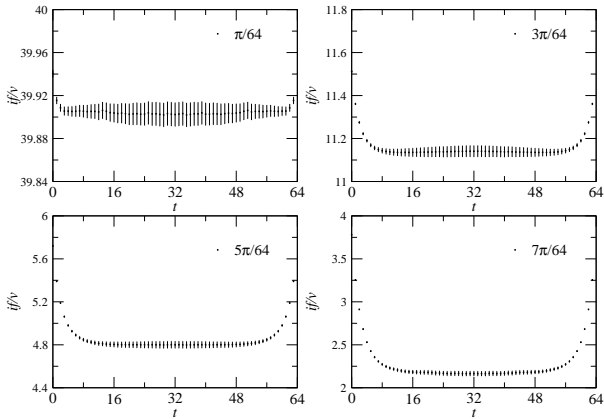}
\caption{\label{fig:dwffdist} Typical plots of the string form factor for $\beta=0.23$, $L=64$, as a function of distance for several $k$. The error bars here are bootstrap estimators for the random error in the correlator measurement; they are therefore correlated at different values of $t$. Only the results for $ 16 \lesssim t\lesssim 48$ are independent of separation and it is results in that region that are used in the other plots presented later in this paper.}
\end{centering}
\end{figure}

The previous section presented evidence to support our approximation that the correlator method we employ in this paper picks up only the ground state of the string. We proceed now to measure the form factor and interpret this as a way of measuring the interactions of the string on various length scales.

The string moves non-relativistically $k \ll \sigma L$ in almost all the cases studied, so while the expressions derived in Section \ref{sec:correlators} allow for this, there is no significant dependence on the form factor with time separation; there are only short-distance effects that could, in principle, be removed.

Let us first demonstrate the applicability of Eq.~(\ref{equ:finalcorr}). Figure~\ref{fig:dwffdist} shows the form of the processed form factor results (we also tested the technique with $T=4L$ and quantitatively identical results were obtained, albeit at considerably greater computational cost). It is assumed that the form factor is pure imaginary, we have no way of determining its phase. We cannot subtract the bulk scalar contribution; this is the principal source of deviations from the expected behaviour at short and intermediate distances.

\begin{figure}[t]
\begin{centering}
\includegraphics[trim=0.5cm 0.5cm 0.5cm 0.5cm,clip=true,scale=0.333,angle=270]{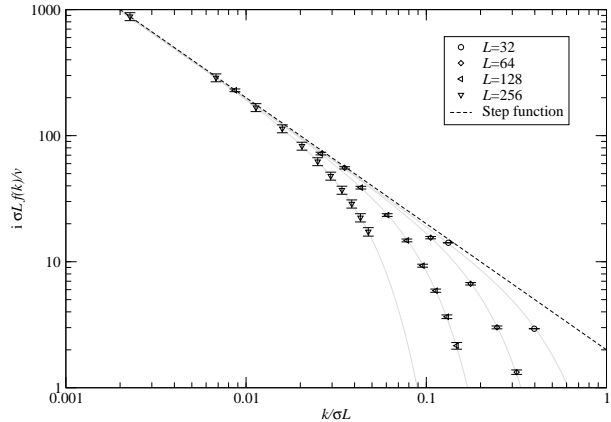}
\caption{\label{fig:dwfffit} String form factor for $\beta=0.23$ and various $L$, with fits given by the ansatz in Eq.~(\ref{eq:broadfitansatz}), assuming the short-distance behaviour takes the form $\tilde{\phi}(k) = 2/k$, implying that the string has no discernible intrinsic width. In this plot and those that follow, the systematic measurement error in $\sigma$ is not shown on the $x$-axis.}
\end{centering}
\end{figure}

Taking results at long distance from correlators similar to those shown in Figure~\ref{fig:dwffdist}, we arrive at Figure~\ref{fig:dwfffit}. Here, appropriately scaled form factor measurements are plotted as a function of the dimensionless combination $k/\sigma L$. For comparison, the step-function result from Eq.~(\ref{eq:ftstepfn}) is shown. A thin string of this form would be independent of volume $L$, and it is clear that at long distances (small $k$), the behaviour of the form factors is independent of volume.

\begin{figure}[t]
\begin{centering}
\includegraphics[trim=0.5cm 0.5cm 0.5cm 0.5cm,clip=true,scale=0.333,angle=270]{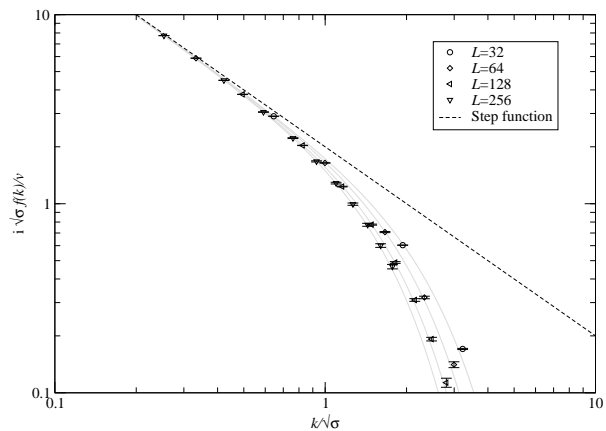}
\caption{\label{fig:notcollapse} As for for Figure~\ref{fig:dwfffit} but scaled instead by $\sqrt{\sigma}$. One would expect the combination $k/\sqrt{\sigma}$ to parameterise the intrinsic width of a string. The data do not collapse onto a single curve.}
\end{centering}
\end{figure}

While Figure~\ref{fig:dwfffit} is scaled in such a way that the results can be compared with the form factors of pointlike solitons (as the $x$-axis plots total momentum over rest energy) and to demonstrate that the string moves non-relativistically in most of the results presented here, to see whether the string form factor were independent of volume a different scaling combination is required. Without logarithmic broadening, we would expect that the data points in Figure~\ref{fig:notcollapse} would collapse onto a single line for all $k$. This is clearly not the case, and we instead have dependence on volume. Irrespective of the scaling employed we see that for larger $L$, the form factor falls off \textsl{faster} with increasing $k$, suggesting that the string indeed becomes broader at larger volumes.

\begin{figure}[t]
\begin{centering}
\includegraphics[trim=0.5cm 0.5cm 0.5cm 0.5cm,clip=true,scale=0.333,angle=270]{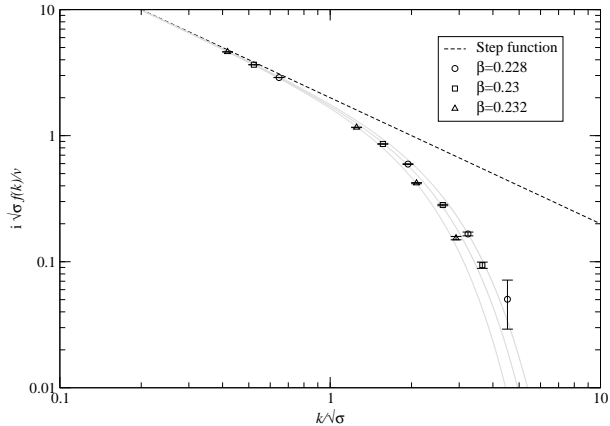}
\caption{\label{fig:varybeta} As for Figure~\ref{fig:notcollapse} but for fixed $L=48$ and various $\beta$. As can be seen, the data do not collapse onto a single line even for constant $L$ due to the weak dependence of $w^2$ on the bulk correlation length. This supports our ansatz, Eq.~(\ref{eq:broadfitansatz}).}
\end{centering}
\end{figure}

Finally, Figure~\ref{fig:varybeta} shows that because $w^2$ (as defined in Section~\ref{sec:dwist}) actually depends logarithmically on the correlation length $\xi$, there is no way to collapse the data onto a single curve.

\subsection{Broadening}

The analytic approximation in Eq.~(\ref{eq:analyticf}) suggests that the string form factor should decay as
\begin{equation}
\label{eq:broadfitansatz}
|f(k)| \approx |\tilde{\phi}(k)| e^{-\frac{1}{2}k^2 w^2},
\end{equation}
where $w^2$ is the `width' of the string due to fluctuations as derived in Section~\ref{sec:fluc}, and $\tilde{\phi}(k)$ is the intrinsic form factor of the unbroadened string. Unfortunately, when the string is moving non-relativistically, it is difficult to distinguish between different choices of $\tilde{\phi}(k)$ -- all of which must fall off as $2iv/k$ at small $k$. For the time being we shall assume that the simplest model -- that of a step function at short distances, given in Eq.~(\ref{eq:stepfn}) -- is valid and take $\tilde{\phi}(k) \propto 1/k$.

\begin{figure}[t]
\begin{centering}
\includegraphics[trim=0.5cm 0.5cm 0.5cm 0.5cm,clip=true,scale=0.333,angle=270]{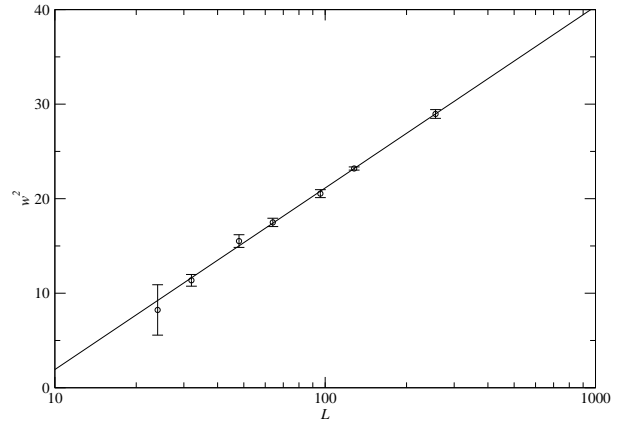}
\caption{\label{fig:broaden}  The values of $w^2$ obtained when fitting Eq.~(\ref{eq:broadfitansatz}) to the data shown in Figure~\ref{fig:dwfffit}; this shows the width of a string as a function of $L$ for $\beta=0.23$. A logarithmic dependence is apparent and a fit of these points to Eq.~(\ref{eq:flucwidth}) is shown, yielding $\sigma=0.0191 \pm 0.0003$ -- a result that is fairly close to but not in excellent agreement with the literature, perhaps due to unaccounted systematic effects. Nonetheless, the behaviour is exactly that of Eq.~(\ref{eq:logwidth}).}
\end{centering}
\end{figure}

If the string had intrinsic width and took the form of a smooth kink then $\tilde{\phi}(k)$ would be given by Eq.~(\ref{eq:ffsemiclass}). However, for $k/(\sigma L) \ll 1$ at fixed $L$ the difference in the two curves is insignificant and therefore we favour the simpler fit arising from step-function model, even if it does not capture all the physics on the shorter length scales. We shall discuss the prospects for measuring an intrinsic width in the conclusions.

It is not possible to establish conclusively based on Figure~\ref{fig:dwfffit} that logarithmic broadening is taking place; indeed the high-$k$ behaviour of Eq.~(\ref{eq:ffsemiclass}) is rather similar. However, in Figures~\ref{fig:notcollapse}~and~\ref{fig:varybeta} we can see that even when correctly scaled the data do not collapse onto a single line for different $\beta$ or $L$.

Finally, the results of fitting Eq.~(\ref{eq:broadfitansatz}) to the data shown in Figures~\ref{fig:dwfffit}~or~\ref{fig:notcollapse} are shown in Figure~\ref{fig:broaden} (although additional values of $L$ are shown in the latter). There is a clear linear relationship between $w^2$ and $\log L$, consistent with Equations~(\ref{eq:logwidth}) and~(\ref{eq:broadenedcorrel}). We must therefore conclude that the behaviour observed is entirely due to the fluctuations in the string.

\section{Conclusions}

We have demonstrated how to study the properties of topological strings and domain walls in quantum field theory using two-point field correlation functions. In the actual numerical calculation we used the 3D Ising model which is in the same universality class as the 2+1D real scalar field theory, and therefore has the same behaviour near the critical point. The model is also exactly dual to the confining $\mathbb{Z}_2$ gauge theory. 

The measured ground state energy $E_0$ shows finite-size behaviour which is compatible with the L\"uscher term generated by quantum fluctuations of Goldstone modes. In the string form factor, these same Goldstone modes suppress short-wavelength modes leading to behaviour that can be interpreted as logarithmic broadening of the effective string width. This is consistent with previous studies of other measures of string widths or domain wall widths.
However, we stress that in contrast with the quantities studied before, the string form factor is a well-defined quantum observable.

Although, in principle, it would be possible to use the form factor to probe the intrinsic structure of the string, this would require excellent data and may not be feasible. Any deviations from zero intrinsic width in the results presented here were within the errors of our fits..

For zero-dimensional topological solitons there are no massless fluctuations and the spectrum of Goldstone modes is relatively straightforward. For example, in the case of a $\lambda\phi^4$ kink, the only Goldstone mode is that which we exploit in measuring the kink's mass. In the present model, this situation is complicated considerably. However we have successfully demonstrated that our technique is again of use.

We would expect our approach to be applicable, with some modification, to confining strings in $\mathrm{SU}(N)$ gauge theory.

\acknowledgements
Our simulations made use of facilities at the Finnish Centre for Scientific Computing CSC. AR was supported by STFC grant ST/J000353/1; KR acknowledges support from the Academy of Finland project 1134018. This work was facilitated by Royal Society International Joint Project JP100273.

\bibliography{ising}

\end{document}